\documentclass[pra,preprint,floatfix]{revtex4-1}

\usepackage{graphicx}
\usepackage{amsmath} 
\begin{document}

\title{Why the Many-Worlds Interpretation?}

\author{Lev Vaidman}
\affiliation{Raymond and Beverly Sackler School of Physics and Astronomy\\
 Tel-Aviv University, Tel-Aviv 69978, Israel}

\begin{abstract}
    A brief (subjective)  description of the state of the art of the many-worlds interpretation of quantum mechanics (MWI) is presented.  It is argued that the MWI is the only interpretation which removes action at a distance and randomness from quantum theory. Limitations of the MWI regarding questions of probability which can be legitimately asked are specified. The ontological picture of the MWI as a theory of the universal wave function decomposed into a superposition of world wave functions, the important parts of which are defined in three-dimensional space, is presented from the point of  view of our particular branch. Some speculations about misconceptions,  which apparently prevent the MWI to be in the consensus, are mentioned.
      \end{abstract}
\maketitle

\section{Introduction}
This is a preface to the Special Issue of {\it Quantum Reports} devoted to the results of the  workshop ``The Many-Worlds Interpretation of Quantum Mechanics:
Current Status and Relation to Other Interpretations,'' Tel Aviv, 18-24 October 2022. 
In my  research of this subject \cite{schizo90,PSA1994,schizo,myMWI,MWI-timesym,myPro,grois,meQMDet,MWIBEll,allpsi,McVa,onto,BornRule,commonsense,WFR}  I find the Many-worlds interpretation (MWI) by far the best interpretation of quantum mechanics. For me, the goal of this workshop is to sharpen the MWI by  reaching a consensus among  supporters of Everett's original idea \cite{Everett} about what the MWI is. 
However, as a scientist,  I must always be sceptical about my beliefs, so I would also consider the workshop as a success if it will demonstrate weaknesses of the MWI
and show  why the MWI should not be accepted as a leading interpretation. Of course, I hope that the result will be different, that we will reach an understanding that the reason for the MWI not  being  the consensus is a mistake in the evolution of science due to a long period of observing quantum phenomena without a satisfactory explanation. This apparently led Bohr to persuade the physics community that quantum mechanics can be used, but cannot be understood, the  statement which until today is frequently made in university courses of quantum theory.

My goal in this paper, which will be available before the workshop, is to set the stage for the workshop: to briefly describe what my version of the MWI is, why I view it as the most preferable interpretation, and what might be  the reasons for misconceptions about the MWI. I invite participants of the workshop (and not only them) to challenge (or improve) my picture in the workshop and in its proceedings. 

\section{
The MWI is the only solution of the measurement problem without action at a distance}

Let me state here what I view as the measurement problem. Today we build single photon sources and single photon detectors and quantum physics explains well the process of photon emission and detection by describing the wave packet of the photon and the particular wave pattern of the ingredients of the single photon detector, Fig. 1.a.
If we add a beam splitter and another detector, Fig. 1.b, the equations of quantum mechanics provide similar wave patterns in the two detectors, $A$ and $B$. Nevertheless, we never observe two simultaneous detections of a single photon by two detectors. This tension between the empirical evidence (single detector clicks) and the physical picture (both detectors change their quantum states) is the measurement problem.

\begin{figure}
\centerline{\includegraphics[width=10.5 cm]{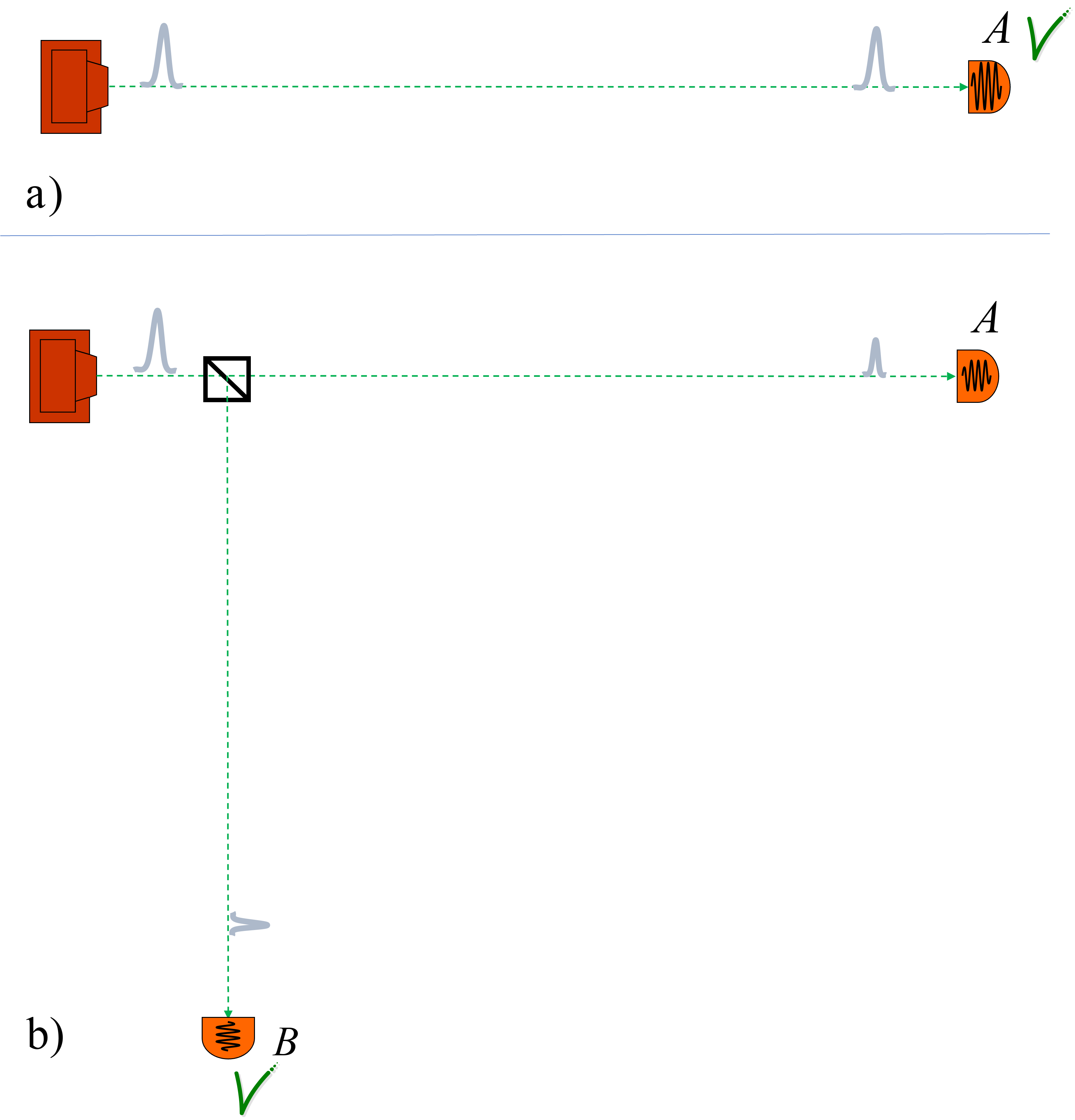}}
\caption{{\bf Measurement problem. } a) The detection of a single photon is fully understood by the creation of a particular quantum wave of parts of the single-photon detector.
b). In the experiment with a single-photon source, beamsplitter, and two detectors, the quantum mechanical equations show a similar (although reduced)  change in two detectors. Nevertheless, we never observe simultaneous clicks of the two detectors.\label{fig1}}
\end{figure}   

Consider now repeating the experiment with the beam splitter without  detectors, see Fig. 2. At the time when the  wave packet is present in place $B$, everyone, independently of their preferred interpretation of quantum mechanics, should agree about the following description of place $B$. Everyone (at least everyone who participates in zero-sum games), should be ready to pay half a dollar for a game in which they get a dollar if a detector placed in $B$  finds the photon. The probability one half of a detection in $B$ is not an ignorance probability, we know everything relevant, but still we bet based on $p=0.5$. (In Bohmian mechanics it is postulated that in the described experiment the Bohmian position cannot be known.)
But now, by  placing a detector shortly before location $A$, we  change the reality in $B$ to probability 0 or 1. An agent near location $A$ will bet with me on the measurement in  $B$ if I do not place the detector before $A$. My action in $A$ can change the  behavior of the agent. The betting behaviour is changed in $A$, but since it is the bet about a measurement in $B$, we witness a superluminal change in $B$.

\begin{figure} 
\centerline{\includegraphics[width=12 cm]{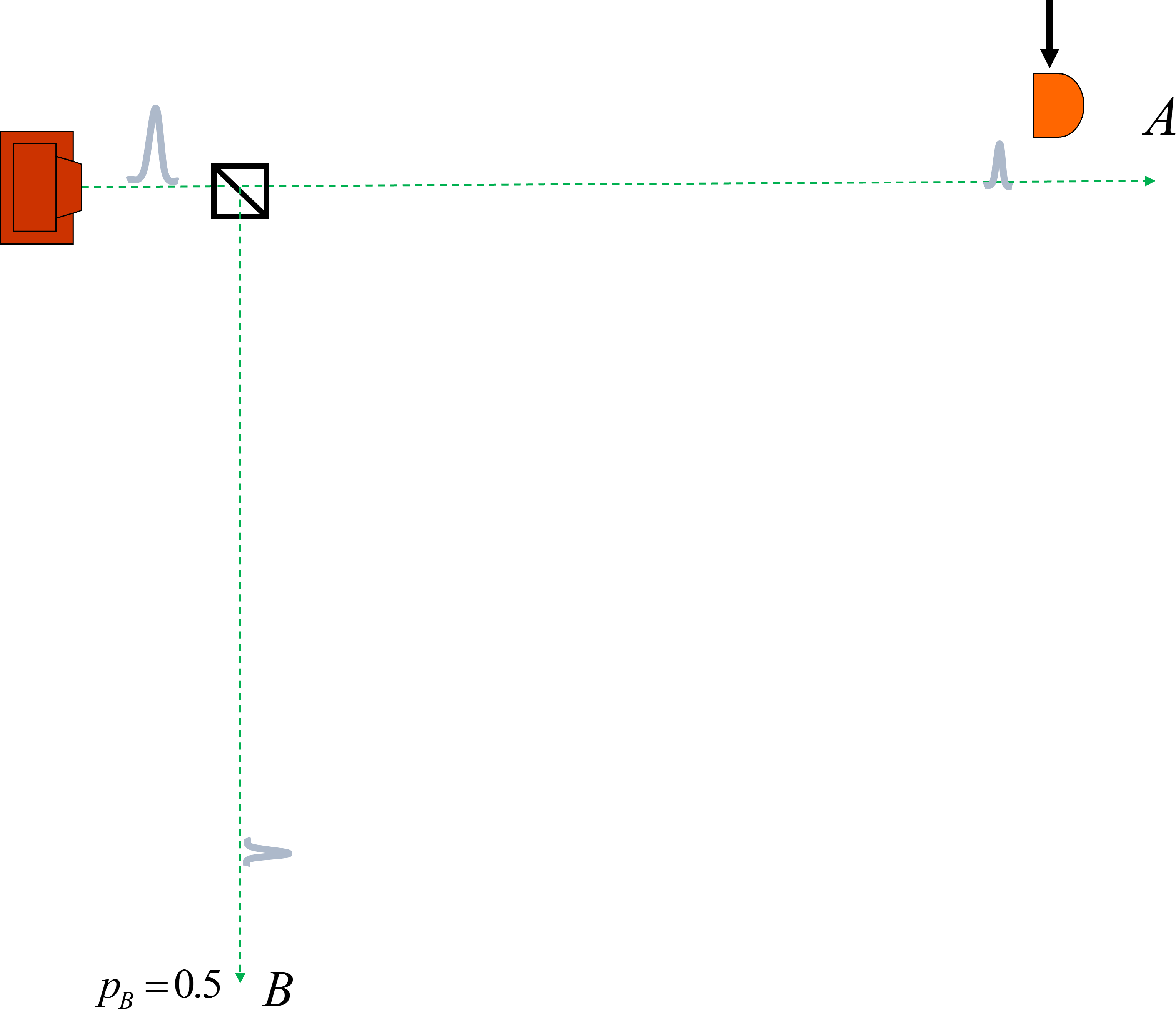}}
\caption{{\bf Action at a distance in a single-world universe. } If we do nothing at $A$, then at a particular moment there will be probability $p=0.5$ of finding a photon at a spacelike separated region $B$. Introducing a detector just before $A$ will lead to a superluminal change in $B$ to $p=0$ or $p=1$.  The change will not be known immediately at $B$, but it does not change the fact that something in $B$ changed, e.g. the readiness of an agent in $A$ to bet about the result of an experiment in $B$.}
\end{figure}   

This argument holds only  in the framework of a single-world interpretation. A believer in the MWI  witnesses the same   change, but it represents the superluminal change only in {\it her} world, not in the physical universe which includes all worlds together, the world with probability 0 and the world with probability 1. Thus, only the MWI avoids action at a distance in the physical universe. The MWI provides a covariant description of the universe made out of quantum particles which allows generalisation to field theory, etc.

\section{
The MWI is the most economical quantum theory regarding the theory's laws}

The title of Everett's thesis ``The Theory of the Universal Wave Function'' \cite{Everett1956} is a good description of the MWI. In this theory, there are no sophisticated collapse mechanisms, no ontology and equation of Bohmian positions, no ``consistent'' or ``decoherent'' histories, no algebras of observables,  no ``relational'' properties with ontological meaning. I   consider the wave function and the Hamiltonian,   responsible for the evolution of the universal wave function, as the only  fundamental entities of the theory, attaching only secondary importance to other operators (observables) by postulating that our experiences supervene directly on the world wave functions. 

 The only  part of our experience, which unitary evolution of the universal wave function does not explain, is the statistics of the results of quantum experiments we performed. We must add a postulate about the probability of self-location in a world which is a counterpart of the Born rule of the standard interpretation \cite{BornRule}.
  Although the self-location  probability postulate explains the observed statistics, it does so without introducing objective chance in Nature: the postulate quantifies an ignorance probability. Thus, the MWI brings back determinism to scientific description \cite{meQMDet}. (Before the quantum revolution, determinism was considered as a virtue of scientific explanation.) We, as agents capable of experiencing only a single world, have an illusion of randomness. This illusion is explained by a deterministic theory of the universe which includes all worlds together.

\section{
The  paradoxes of the quantum theory are resolved in the framework of the MWI interpretation }
The MWI provides simple answers to almost all quantum paradoxes. Schr\"{o}dinger's Cat is absurd in one world, but unproblematic when it represents one world with a live cat and a multitude of worlds with the cat which died at different times of detection of the radioactive decay.

It is very unfortunate that we do not know what would be the reaction of Einstein to the MWI. It seems that he would adopt it, because it resolves two main difficulties Einstein had with quantum mechanics: randomness and action at a distance. 

The paradoxical behaviour of Bell-type experiments disappears when quantum measurement does not have a single outcome \cite{MWIBEll}. Since the spin measurement of one particle of an Einstein-Podolsky-Rosen pair  has both results Up and Down, the physical description of the second particle as a mixed state is not changed at the moment of the measurement of the first particle at a space-like separate location.

The paradoxes in describing collapse in different Lorentz frames \cite{AA84} do not arise when the theory does not have  collapse. The paradox of interaction-free bomb testing \cite{EV} in which we get information about a region without a probe being there, is resolved by interaction of the probe in a parallel world. The paradox of the amount of information transferred in teleportation is resolved by the nonlocality of worlds and an observation that the only information remaining to be transferred after the local Bell measurement   is the identity of the world we are in \cite{PSA1994}. Finally, recent paradoxes appearing in the description of pre and postselected quantum systems: the 3-box paradox \cite{AV91}, Hardy paradox \cite{Hardy+}, the quantum pigeon holes conundrum \cite{pigeon}, and discontinuous traces in nested interferometers \cite{past} are all resolved by the fact that there are parallel worlds with different postselections \cite{commonsense}.

\section{
Conceptual changes in our approach to a  scientific theory that should be made when we accept the world splitting structure of the universe }

Up until a particular point in time (our present) there is no difference in our experience  between the single world of the universe in which quantum mechanics includes collapses at every quantum measurement and the corresponding world of the MWI universe. In Fig. 3a the whole tree of worlds of the MWI is schematically shown, in Fig. 3b  our world until the present and all our future worlds, and in Fig. 3c the corresponding world of the universe with collapsing worlds is shown.   There is no difference in the description of the past between the MWI and the theory with collapse at every measurement. However, there is a difference for the future. While in the collapsing universe there is a diachronic identity of the world towards the past and future, in the MWI, there is no diachronic identity towards the future.

\begin{figure}
\centerline{\includegraphics[width=15 cm]{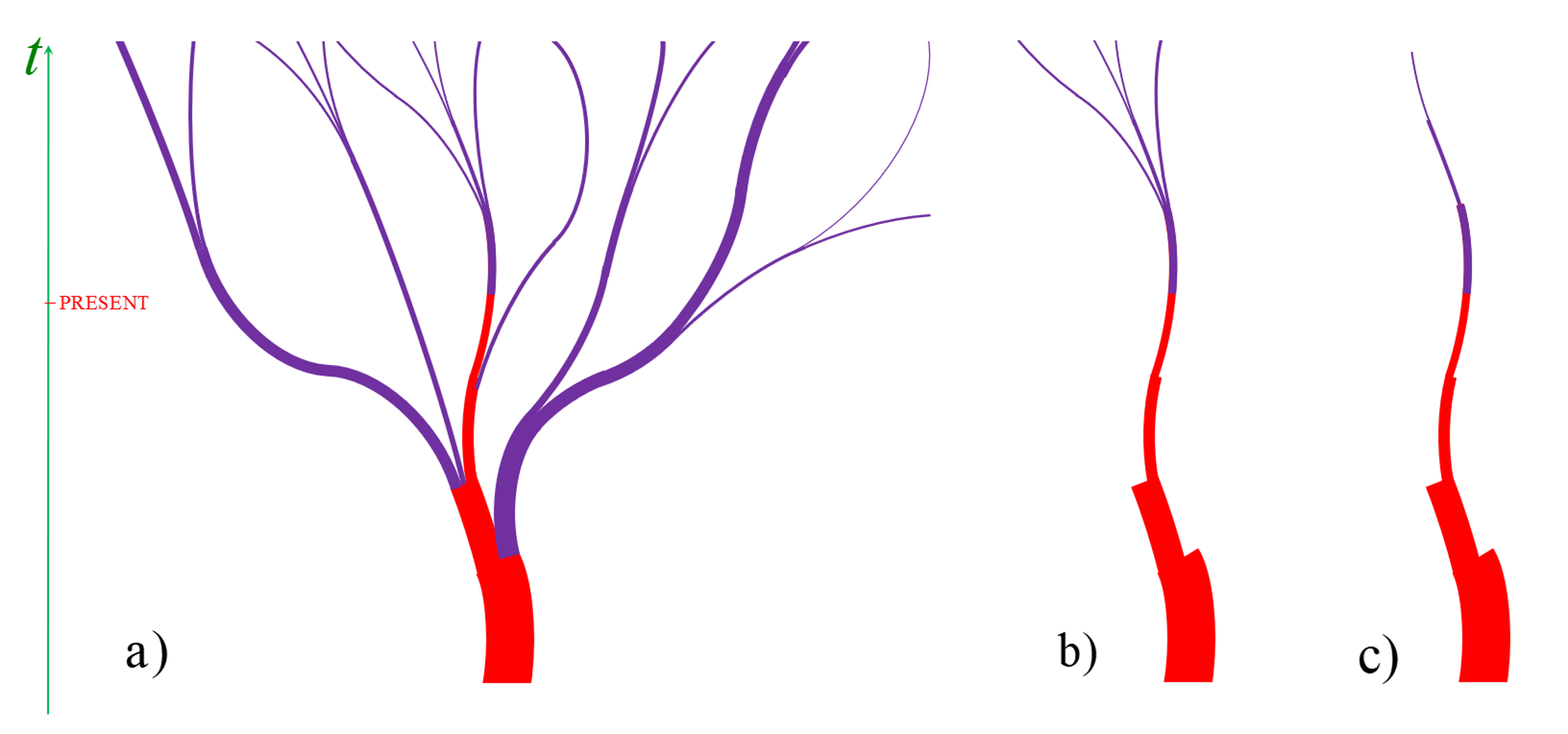}}
\caption{{\bf The world structure in the MWI and  a  single-world universe. } a) The whole tree of many worlds in the MWI. b) One world of the MWI until present together with the tree of future worlds splitting out of it in the future. c) The corresponding world of the theory with collapse.}
\end{figure}   

The  difference in the world splitting structure of our universe should be reflected in our attitude towards the past and future. In our memories there is a single world. In this world, in our past, we can identify deterministic as well as chancy (in the case of observing results of quantum measurements) events. We understand that chancy events are our illusion in a deterministic  physical universe due to  our construction which does not allow the experience of superpositions.  We understand the existence of parallel worlds in the past, but our memories define a unique diachronic identity over the past. By contrast, we do not have diachronic identity in the future. There are multiple worlds (created by future quantum measurements) which are {\it all} related to our world at present, Fig. 3b. Thus, we cannot ask what is the probability for a  result of a quantum measurement to be performed. It should not prevent us to behave in a ``normal'' way (as believers of a single collapsing world picture). The justification is very different:
we care for all future parallel worlds according to their ``measure of existence'' \cite{schizo,grois} which is proportional to the objective probabilities of the corresponding possible collapsing worlds.

\section{
What is a ``world'' in the MWI? }

The ``world'' in my MWI  is not  a physical entity. It is a 
 a term defined by us (sentient beings) which helps to connect  our experience with the ontology of the theory, the universal wave function.
My definition \cite{myMWI} is: 
\begin{quote}
 {\it
A world is the totality of macroscopic objects: stars, cities, people, grains of sand, etc. in a definite classically described state.}   
\end{quote}
Just as our experience is vague, so a world is vaguely defined: ``macroscopic'', ``classically definite'' are not rigorous terms. For a particular choice of these terms,``world'' has a physical counterpart as the world  wave function which is a part of the superposition of the universal wave function. Until the next splitting, it autonomously evolves, but contrary to a popular view \cite{MWI2010}, it has nothing to do with {\it decoherence due to the environment}. The world wave functions of different worlds do not interfere, unless super-technology {\`a} la Wigner with his friend is present, because in real situations we cannot arrange interference of macroscopic bodies (even if another super-technology  switches off the decoherence with the environment).

The MWI is ``The Theory of the Universal Wave Function'', but the starting point in our description is our world, not the universal wave function. The ``emergence'' program \cite{wallacebook} is not simple, and it is also not needed. In any case, we have very little information about the universal wave function, so the emergence program, even if successful, is of little practical value.

We do know a lot about our world.  There is no question of a preferred basis, it is defined by our world, see Fig. 3b. Every physicist who does not worry about the interpretation and accepts the von Neumann process I happening at every quantum measurement, has no difficulty to describe the basis in which she observes our single world (so she believes), and the MWI believer uses the same basis.  
It is an obviously correct statement that the world of a dead cat is stable, while the world ``plus'', the plus superposition of dead cat and alive cat, in almost no time evolves into  equal weight superposition of the worlds ``plus'' and ``minus''. However, in my framework there is no need to analyse this, because the worlds I define have no cats in a superposition.

The MWI believer, being aware of recent quantum measurements, has information about some  parallel worlds. It is easier for her to think about coherent splitting in a quantum computing device \cite{deutsch1992rapid}, but the main reason for introducing parallel worlds is to avoid collapse (the von Neumann process I). This  makes physics elegant, deterministic, and without action at a distance.

\section{Connection between our experience and the universal wave function
 } 

A popular question is: what is the space in which we consider the world wave function: is it a configuration space or three-dimensional space \cite{albert2013,Maudlin}? The answer is a subtle one \cite{WFR}: the macroscopic objects are well localised and are not entangled within the world wave function, so every macroscopic object is represented by a product of the wave function of some collective variables defined in three dimensions,  times the entangled state in the configuration space of degrees of freedom of the microscopic parts of the object. 

The theory of our brain is not developed enough, but the hope is that the wave function of some collective variables of its constituents in three-dimensional space directly corresponds to our experience. To avoid dealing with brain science, it is reasonable to assume that our senses faithfully observe the three-dimensional picture of macroscopic objects. Then,  the  three-dimensional wave function of the collective variables of macroscopic objects is the bridge between the world wave function and our experience.

Instead of collective variables with the wave function in three dimensions one can consider the spread of the world wave function in three dimensions, which is very similar to the ``mass density'', the primitive ontology or ``local beables'' of alternative approaches \cite{Allori2014}. However, I do not see the necessity to add an ontic status to the mass density, it is included in the ontology of the world wave function  which also allows more efficient ways of describing objects, e.g. the three-dimensional density of organic molecules for getting a more precise picture of living organisms. In particular, when such a three-dimensional density looks like me, I postulate that this construction ``experiences'' my feelings.

\section{The (illusion of) probability in the  MWI}

The MWI is a deterministic theory, but the determinism is manifested on the level of all worlds together. This is the level of a mathematically rigorous physical theory. We live (or more precisely, lived) in one world with  random probabilistic events (results of quantum measurements). Indeed, the complete knowledge of the wave function of our world, prior to a quantum measurement, does not specify a particular outcome. Usually, the outcome cannot be presented as uncertain due to ignorance of details in the measurement setup. The only way to introduce ignorance is to apply a ``sleeping pill'' trick \cite{schizo} which leads to a situation in which an observer splits according to the outcome of the measurement without being aware of the outcome. Then she (and only she!) is ignorant about what is the world she lives in. The observer does not have a concept of probability of an outcome (she knows that all possible outcomes of the experiment take place), but she has a legitimate concept of probability of  self-location in a world with a particular outcome.

A separate issue is the quantitative question: what is the probability of self-location in a particular world? I claim that it has to be postulated in addition to the postulate of unitary evolution of the universal wave function and a postulate of the correspondence between the three-dimensional wave function of an observer within a branch and the experience of the observer. The postulate is that the probability of self-location is proportional to the ``measure of existence'' \cite{schizo,grois} which is a counterpart of the Born rule of the collapse theories. 

Apart from empirical evidence, there are many natural principles which, together with symmetry considerations, suggest the plausibility of the self-location rule. For example, it is enough to postulate, that when a quantum measurement performed in one world splits it into several worlds, then the probability of self-location in the first world is equal to the sum of the probabilities of self-location in all the newly created worlds. Never-mind how plausible this or other principles  taken as a basis of the MWI Born rule proof are (I have a proof based on the impossibility of superluminal signalling), some  principle is necessary \cite{BornRule}. The postulate of the unitary evolution of the universal wave function alone is not enough.  Note, that the necessity of the additional postulate in the framework of the MWI is less obvious than in the framework of collapse interpretations, in which the Born rule is clearly a separate postulate  describing a nonunitary process.

\section{What might be the reasons for the MWI not being in a consensus?}

The reluctance of a human to accept the MWI is natural. We would like to think that we are the center of the Universe: that the Sun, together  with other stars, moves around Earth, that our Galaxy is the center of the Universe, and we are unhappy to accept that there are many parallel copies of us which are apparently not less important.

The next issue  is the difficulty to apprehend what exactly a parallel quantum world means. It is misleading to view the universe  as a multitude of (countable) classical worlds  created by a magician.  The cosmological multiverse is  very different and much easier to understand. 

Negative publicity for the MWI  comes from the controversial claims about advantages of the MWI relative to other interpretations, e.g., that the Born Rule can be derived instead of postulated \cite{deutsch1999}. The claim is natural, because it is not simple to postulate the Born Rule in the MWI, but I believe it is false. In any case, the difficulties of this program reflect  negatively on the MWI.

Another source of negative publicity is the controversy generated by presenting MWI as a theory of the universal wave function on  configuration space \cite{albert2013}, obscuring the connection between ontology and our experience. Avoiding non-separability by moving to configuration space \cite{Ney2021}, is hardly helpful.

In my view,  similar damage comes from an attempt to present MWI in the Heisenberg picture with a  controversial claim of bringing separability into quantum mechanics \cite{DeutschHayden}. The Heisenberg picture provides not just a description of the present, but also of the past, so it is nonlocal not only in space, but also in time. Assuming the initial state as given, and describing reality by multiplied local Hilbert spaces which include all systems interacting with  local systems in the past, achieves formal locality including separability, but for the price of enormous complexity  \cite{bedard}.

\section{Conclusions} 
Let me summarise the main points of my approach to the MWI for which I am looking for support/refutation in the upcoming workshop.

a) The lack of action at a distance is a huge physical advantage which is not present in other interpretations.

b) Determinism is a huge philosophical advantage which is not considered as such due to an error in the evolution of science (apparently explained by not seeing a deterministic option for physics for too long).

c) The MWI allows us to view physics in three spatial dimensions within the particular world of the MWI we live in. (But we should not  disregard  nonlocality of entanglement which requires the configuration space for its description.)

d) Our world defines our world wave function (the alleged preferred basis problem) and the difficult emergence program does not need a solution.

e) There is only an illusion of probability of outcomes of quantum measurements. It naturally leads to an effective Born Rule via measures of existence of  worlds (and can be given an ignorance probability meaning as the probability of self-location in a particular world). Quantum worlds, contrary to classical worlds, might have measures of existence which are not just zero or one.

This work has been supported in part by the Israel Science Foundation Grant No. 2064/19.

\bibliography{reference}

\end{document}